%% file: catalystPore_main.tex
\documentclass{oupau} 

\usepackage{latexsym, amssymb, enumerate, amsmath}
\usepackage{hyperref}
\usepackage{color}
\usepackage{tikz}
\usepackage{graphicx}
\usetikzlibrary{decorations.pathreplacing}
\usetikzlibrary{shapes,decorations,shadows}
\usetikzlibrary{decorations.pathmorphing}
\usetikzlibrary{decorations.shapes}
\usetikzlibrary{fadings}
\usetikzlibrary{patterns}
\usetikzlibrary{calc}
\usetikzlibrary{decorations.text}
\usetikzlibrary{decorations.footprints}
\usetikzlibrary{decorations.fractals}
\usetikzlibrary{shapes.gates.logic.IEC}
\usetikzlibrary{shapes.gates.logic.US}
\usetikzlibrary{fit,chains}
\usetikzlibrary{positioning}
\usepgflibrary{shapes}
\usetikzlibrary{scopes}
\usepackage{chemfig}

\definecolor{dgreen}{rgb}{0,.5,0}

\definecolor{grau}{gray}{.5}
\definecolor{schwarz}{gray}{0}

\input{user_comm.tex}

\newtheorem{lem}{Lemma}
\newtheorem{rem}{Remark}

\DeclareGraphicsExtensions{.png,.eps,.ps}
\graphicspath{{archive_pics/}}

\begin{document}

\runningheads{M. Schmuck and P. Berg}{Effective macroscopic catalyst layer model}
\title{Homogenization of a catalyst layer model for periodically distributed pore geometries in PEM fuel 
cells}

\author{Markus Schmuck\affil{a}\corrauth\\
Peter Berg\affil{b}}

\address{\affilnum{b}Faculty of Science, UOIT, Canada, 2000 Simcoe St. N., Oshawa, ON, L1H 7K4. E-mail: peter.berg@uoit.ca}
\corraddr{Departments of Chemical Engineering and Mathematics, Imperial College, London SW7 2AZ, UK. E-mail: m.schmuck@imperial.ac.uk}

\begin{abstract} We formally derive an effective catalyst layer model comprising the reduction of oxygen for periodically distributed pore geometries. By assumption, the pores are completely filled with water and the surrounding walls consist of catalyst particles which are attached to an electron conducting microstructure. The macroscopic transport equations are established by a multi-scale approach, based on microscopic phenomena at the pore level, and serve as a first step toward future optimization of catalyst layer designs.
\end{abstract}

\keywords{mutli-scale analysis, Butler-Volmer reactions, upscaling, homogenization, thin-double-layer limit}
\smallskip

\received{\date}

\maketitle

\section{Introduction}\label{sec:Intro}
\input{realIntro.tex}

\section{Problem description}\label{sec:Pr}
For the presentation of the main results, we first need to explain the problem of cathode catalyst layers in 
proton exchange membrane fuel cells (PEMFC) at the level of a single pore 
in Section \ref{sec:SiPoMo} and then extend it to the whole catalyst 
layer described by a periodic representation of such single pores in Section \ref{sec:MoRe}.

\subsection{Single pore model}\label{sec:SiPoMo}
\input{intro.tex}

\section{Main results}\label{sec:MaRe}
\input{mainres.tex}

\section{Formal derivation of the upscaled system \reff{EfEqs} by the multiple-scale method}\label{sec:MaDe}
We assume that the periodic formulation \reff{PeCaLa}--\reff{Ir} is well posed and immediately derive 
the homogenized problem via the multiple-scale method.
 \input{multiScale.tex}

 \section{Application to straight channels}\label{sec:Ap}
\input{applic.tex}
\medskip

\ack{The first author acknowledges the support by the Swiss National Science Foundation (SNSF) through 
the grant PBSKP2-12459/1 during his stay at MIT and the second author the support by an NSERC Discovery Grant.}

\medskip

          \bibliographystyle{plain}  
          \bibliography{catalystPoreLayer}

\end{document}

%% file: user_comm.tex
\newcommand{\reff}[1]{(\ref{#1})}
\newcommand{\ol}[1]{\overline{#1}}

\newcommand{\Div}{\textrm{div}\,}

\newcommand{\x}{{\bf x}}
\newcommand{\y}{{\bf y}}

\newcommand{\cg}[1]{\mathcal{#1}}

\newcommand{\av}[1]{\left|#1\right|}

\newcommand{\N}[2]{\left\|#1\right\|_{#2}}
\newcommand{\brkts}[1]{\left(#1\right)}
\newcommand{\ebrkts}[1]{\left[#1\right]}

\newcommand{\brcs}[1]{\left\{#1\right\}}

\newcommand{\bsplitl}[2]{
\begin{equation}
\begin{split}
#1
\end{split}
\label{#2}
\end{equation}}


%% file: realIntro.tex
The global economy is faced with a transition towards a renewable energy infrastructure. The optimization of energy efficient devices and the development of new and environmentally friendly strategies for the production of electrical energy are key issues in this context. One promising field in this direction is fuel 
cell research. In particular, polymer electrolyte fuel cells might become future power sources for 
portable devices and motorized transport because of their high energy densities and thermodynamic efficiency. 

However, a major proportion of voltage losses occurs in the cathode catalyst layer (CL)
\cite{Chan2011}, thereby lowering the efficiency of the cell. High platinum (Pt) loadings are required to enhance the oxygen reduction reaction 
such that sufficient current densities are achieved. 
The long-term target is to reduce current platinum use by an order of magnitude.
Consequently, optimization of CL design so as to reduce voltage losses and the amount of costly platinum is a central goal in polymer electrolyte membrane (PEM) fuel cell research, see \cite{AprDOEHP2008}.

The catalyst layer in PEM fuel cells is comprised of a complex multiphase
porous structure. Normally, it is a three-phase random composite of open gas pores and carbon-supported 
Pt in which ionomer strands are embedded. Nanometer-size platinum particles, the catalyst, are supported on larger carbon (C) agglomerate particles which are excellent electronic conductors. Proton transport occurs from the anode towards the cathode catalyst layer through the PEM and ultimately, inside the CL, through the ionomer which consists 
of nano-thin pieces of polymer electrolyte membrane. Oxygen enters through the gas diffusion layer (GDL) and diffuses into the catalyst layer inside the gas phase, where the oxygen concentration decreases toward the catalyst particles (Pt). Electrochemical reactions preferentially occur at the interface between the catalyst particles and the electrolyte (i.e. the ionomer). Only catalyst particles that are simultaneously accessible to electrons, protons, and oxygen are electrochemically active. Strong non-equilibrium effects for gas-phase flows in small-scale confined geometries as well as liquid water generated from the electrochemical reaction
\bsplitl{
\frac{1}{2}{\rm O}_2 +2{\rm H}^+ + 2e^-
	\quad\rightleftharpoons\quad {\rm H}_2{\rm O}\,,
}{R}
complicate an exact description of proton transport in the catalyst layer.

So far, little is known about the influence of geometrical parameters such as 
porosity, pore size, and surface properties of the C-Pt phase on proton transport and macroscopic 
reaction rates. Hence, 
systematic optimization of these parameters requires physical and mathematical understanding of relevant phenomena. The recent developments of ultra-thin catalyst layers (UTCLs) such as 3M's organic perylene whiskers, see for example 
\cite{Debe2006}, carbon nanotubes \cite{Simon2008}, or carbon produced by template techniques where 
the porosity is controlled by a silica matrix \cite{Barsukov2003}, call for models which reliably account for 
the influence of the pore geometry on the transport properties. In addition, these nano-scale geometries can be designed without ionomer. In this case, liquid water facilitates proton transport    
which is what we will study in this contribution.

To date, merely volume averaging is applied to fixed geometries like spheres or cylinders, capitalizing radial symmetry under restricting and simplifying assumptions \cite{Chan2011,Eikerling2008}. Hence, the goal of this article is to provide an upscaled macroscopic 
description which consistently describes general geometries/designs of the catalyst layer with the help of a 
microscopic periodic reference cell. Such a generalization will serve as a promising extension for the 
description of the new ultra-thin catalyst layers, carbon nanotubes and general meso-porous materials 
designed by template techniques.

In general, volume averaging approaches cannot treat nonlinear models 
and require the choice of an appropriate test volume whose size is not obvious. As a consequence, such averages are less rigorous 
than homogenization techniques like the two-scale convergence method, for example \cite{Allaire1992,Nguetseng1989}. Here, we employ formal periodic homogenization 
which is a multi-scale approach, see for example \cite{Bensoussans1978,Cioranescu2000}. 
It allows to formally obtain the effective macroscopic description \reff{MaEfEq} without technical a priori estimates, followed by a convergence analysis. This is a subtle point since the convergence of such reactions impose new difficulties and open questions for a rigorous analysis of such interface phenomena.

From an analytical perspective, it is interesting to note that there exists an electro-osmotic flow problem in cylindrical channels of PEM that can be solved explicitly without numerical tools \cite{Berg2011}. These solutions might be useful for test purposes of numerical approaches. However, such solutions
lack the reaction kinetics at the boundaries of the domain as in catalyst layer pores.

\medskip

Let us briefly summarize the main result of this article. We derive effective macroscopic equations for a general cathode catalyst layer $\Omega\subset\mathbb{R}^d$ containing reactions on its phase interfaces $S^c$, as explained for a cylindrical 
pore in Section \ref{sec:SiPoMo}. The spatial dimension is denoted by $1\leq d\leq 3$. The relevant physical quantities 
(in dimensionless form) are the oxygen concentration $C_O$, the 
proton concentration $C_+$, and the electrostatic potential $\Phi$ while water flux is neglected. However, since no water is produced at the anode, the macroscopic model derived here describes also well anodic currents.
The new contribution of this article is the systematic derivation of reference cell problems which reliably capture the characteristic pore geometry of the considered 
catalyst layer. 
Solutions of these reference cell problems define effective diffusion tensors for oxygen, $\hat{\rm D}^O$, 
for protons, $\hat{\rm D}^+$, an effective mobility tensor, $\hat{\rm M}^+$, and an effective electric permittivity tensor $\hat{\varepsilon}$, see Theorem \ref{thm:EffMo} in Section \ref{sec:MaRe}. The new effective catalyst layer model reads (in dimensionless form) as 
follows
\bsplitl{
(\textrm{\bf Upscaled model:})\quad
\begin{cases}
-{\rm div}\brkts{\hat{\rm D}^O\nabla C_O}
	= \ol{\beta}_O(C_+)^{n_+}(C_O)^{n_O}\exp\brkts{-\alpha_c(\Phi-\Phi_0)}\,,
	&\qquad\textrm{in }\Omega\,,
\\
-{\rm div}\brkts{\hat{\rm D}^+\nabla C_+ 
	+ C_+\hat{\rm M}^+\nabla\Phi}
	= \ol{\beta}_+(C_+)^{n_+}(C_O)^{n_O}\exp\brkts{-\alpha_c(\Phi-\Phi_0)}\,,
	&\qquad\textrm{in }\Omega\,,
\\
-{\rm div}\brkts{\hat{\varepsilon}(\lambda^2,\gamma)\nabla\Phi}
	=pC_++\rho_s\,,
	&\qquad\textrm{in }\Omega\,,
\end{cases}
}{MaEfEq}
where $p$ is the porosity, $\rho_s(x):=\frac{1}{\av{Y}}\int_{I}\sigma_s(x,y)\,do(y)$ represents an effective surface charge on the 
pore walls $I:=\partial Y^1\cap\partial Y^2$ for a given surface charge density $\sigma_s$, $\ol{\beta}_O
	:= \frac{i_0L\Lambda}{4 e {\rm D}_O}$ and $\ol{\beta}_+
	:= \frac{i_0L\Lambda}{e {\rm D}_+}\,,$ are dimensionless numbers mediating the coupling to interfacial 
reactions, $L$ is the characteristic length of the catalyst layer, and the parameters $n_+$ 
and $n_O$ denote reaction orders. $\Lambda$ stands for the Lebesgue measure of the interface, i.e., 
$\Lambda:=\av{I}$, where $I$ is the pore-solid interface constituting the pores walls. The variable $\Phi_0$ denotes the standard (equilibrium) potential. The 
parameter $\lambda:=\frac{\lambda_D}{L}$ represents the dimensionless Debye length 
and $\gamma:=\frac{\epsilon^s}{\epsilon^p}$ is the dimensionless electric permittivity. The constants 
$\lambda_D,\,\epsilon^s$ and $\epsilon^p$ stand for the Debye length $\lambda_D:=\brkts{\frac{\epsilon^pRT}{2z^2_+eF\ol{c}}}^{1/2}$, the electric permittivity of 
the solid and the pore phase, respectively. $\ol{c}$ denotes 
the reference salt concentration. The parameter $z_+$  is the charge number of the protons $C_+$.

The new effective catalyst layer model \reff{MaEfEq} allows for similar limit considerations with respect to 
$\lambda$ and $\gamma$ as initiated in \cite{Schmuck2011}. These 
convenient and advantageous limit properties of \reff{MaEfEq} are due to the generality of the homogenization 
procedure performed here. Since especially the thin-double-layer limit is a widely used approximation for the 
Poisson-Nernst-Planck system in engineering, we are able to compare related results in engineering with the systematically derived macroscopic system \reff{MaEfEq} by mathematical homogenization theory herein. As already stressed in \cite{Schmuck2011}, the effective Nernst-Planck equations agree 
very well with the heuristically suggested models in \cite{dydek2011,He2009,Ramirez2003,Szymczyk2010}. 
We point out that the system \reff{MaEfEq} extends results from previous articles \cite{Schmuck2011,Schmuck2011a} by including Butler-Volmer reactions on the solid/liquid interfaces in the pores. Due to the additional non-linearities arising 
in the coupled system \reff{MaEfEq}, a careful fixed point iteration scheme is required to solve such a problem. Systematic strategies for the development of such schemes can be found in Jerome's book \cite{Jerome1996} for instance.

Several references exist in the literature which heuristically motivate effective (upscaled) equations for the catalyst layer that are sometimes close or similar to the new and mathematically derived formulation \reff{MaEfEq} in here, see for example \cite{Biesheuvel2011,Chan2011}. Hence this article strives to fill this gap with a systematic derivation. A different approach to the derivation in this 
contribution can be found in \cite{Paddison2001} where a non-equilibrium model, using  statistical mechanics, is applied to compute effective friction and diffusion coefficients in Nafion membranes. Our work utilizes such computed coefficients at the pore level for the derivation of the new macroscopic equations \reff{MaEfEq} at the scale of the catalyst layer.

\medskip

Section \ref{sec:SiPoMo} introduces the relevant equations for a 
single pore. In Section \ref{sec:MoRe}, a periodic representation of the catalyst layer provides the microscopic representation which is the starting point for the upscaled results which are presented in Section \ref{sec:MaRe} and derived in Section \ref{sec:MaDe}.

%% file: intro.tex
\begin{figure}
	\centering
\includegraphics[height=6cm,width=11cm]{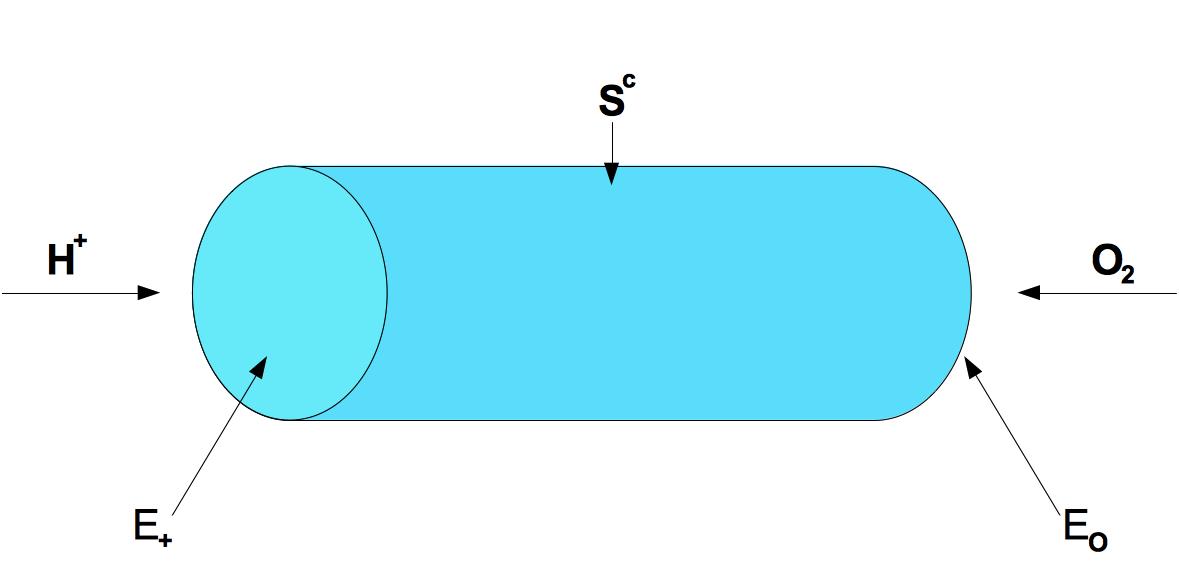} 
	\caption{Single water-filled pore $\Omega^p$ in the cathode catalyst layer: {\bf Left entrance $E_+$:} Flux of protons enters through the polymer electrolyte membrane (PEM). {\bf Cylinder wall 
	$S^c$}: Surface where the reaction takes place. It is the interface between the catalyst and the water-filled pore. {\bf Right entrance $E_O$:} Flux of oxygen enters through the gas diffusion layer (GDL).}
	\label{fig:CaPo}
\end{figure}

We begin with equations describing a single 
water-filled pore. For simplicity, we restrict our analysis to 
the cathode catalyst layer. The analysis for the anode 
layer is then an immediate adaptation of our results.
We first formulate a model for a single pore with boundary conditions defined by the physical fluxes as 
depicted in Figure \ref{fig:CaPo}. We are neglecting the flux of liquid water and assume the pore is uniformly 
filled with water.\\

In order to place our considerations on a well-defined thermodynamic basis, we require that our 
physical quantities of interest, i.e., oxygen concentration $C_O$, proton concentration $C_+$, 
and electric potential $\Phi$, constitute the following dimensionless bulk free energy,
\bsplitl{
H(C_O,C_+,\Phi) =\frac{1}{RT}(U-TS)
	& := \int_\Omega \brkts{
		 \sum_{k\in\brcs{O,+}} C_k\brkts{ {\rm log}\,C_k -1}
		+ C_+\Phi 
		-\lambda^2 \av{\nabla\Phi}^2
		}\,dx\,,
}{HeFrEn} 
where $U$ and $S$ represent the internal energy and the entropic contribution, respectively, and 
$\Phi:=\frac{z_+F\phi}{RT}=\frac{ez_+\phi}{kT}$ is the dimensionless electric potential. The free energy \reff{HeFrEn} allows us to define the chemical 
potentials $\mu_O(C_O)$ and $\mu_+(C_+,\Phi)$ associated with the catalytic pore governed by $C_O$ and $C_+$. These potentials are defined as the 
Fr\'echet derivative of $H$, that means,
\bsplitl{
\mu_O(C_O)
	& := \frac{\delta H}{\delta C_O}
	= {\rm log}\,C_O\,,
\\
\mu_+(C_+,\Phi)
	& := \frac{\delta H}{\delta C_+}
	= {\rm log}\,C_+ +\Phi\,.
}{ChPo}
Motivated by Onsager \cite{Onsager1931}, it is generally accepted that a thermodynamic system is driven by gradients of associated chemical potentials. Hence, we can define the fluxes $J_O$ and $J_+$ corresponding to \reff{ChPo} by
\bsplitl{
J_O
	& := C_Ok_BT{\rm M}^O\nabla\mu_O
	= {\rm D}^O\nabla C_O
	\,,
\\
J_+
	& := C_+k_BT{\rm M}^+\nabla\mu_+
	= {\rm D}^+\brkts{\nabla C_+ +C_+\nabla\Phi}
	\,,
}{Fl}
where we applied Einstein's relation ${\rm M}^k=\frac{D^k}{k_BT}$ for $k\in\brcs{+,O}$ and the identity $\frac{RT}{F}=\frac{kT}{e}$.
We require that oxygen and proton concentrations are conserved quantities and hence satisfy the 
continuity equations,
\bsplitl{
\frac{\partial C_O}{\partial t}
	& = {\rm div}\brkts{J_O}\,,
	\qquad\textrm{and}\qquad
\frac{\partial C_+}{\partial t}
	= {\rm div}\brkts{J_+}\,,
}{CoEq}
where we made use of the diffusion timescale $\tau_D=\frac{\ell^2}{{\rm D}}$ for a characteristic length scale $\ell$.
For simplicity, we continue our considerations with the time-invariant formulations of \reff{CoEq} and 
hence set $\frac{\partial C_O}{\partial t}=\frac{\partial C_+}{\partial t}=0$. We also neglect 
any possible source or sink terms in the bulk of the microscopic formulation. However, a major result of 
this article is the systematic derivation of such source and sink terms in the new upscaled catalyst layer 
model. Finally, we remark that the concentrations $C_O$ and $C_+$ are extended by zero in the solid phase and again denoted by $C_+$ and $C_O$ for notational convenience. This 
physically means that the solid phase is considered as an ideal conductor such that any 
charge accumulation is immediately equilibrated. The electric potential is defined on both solid and liquid phases by taking into account the different electric permittivities. Such a regularization is related to 
a diffuse interface approach and hence physically meaningful. Finally, we account for different electric 
permittivities in the pore and the solid phase by $\hat{\varepsilon}(x):=\lambda^2\chi_{\Omega^p}(x)
+\gamma\chi_{\Omega\setminus\Omega^p}(x)$.

Next, we move our focus from the bulk of the pore to the solid/liquid interface. Along the pore 
wall $S^c$, see Figure \ref{fig:CaPo}, we have to account for 
the reduction reaction of oxygen. This reaction, introduced in \reff{R}, is explained in detail in 
standard literature on electro-chemistry \cite{Newman2004,Plieth2008}.
We now derive a boundary condition on the pore wall related to this reaction.  

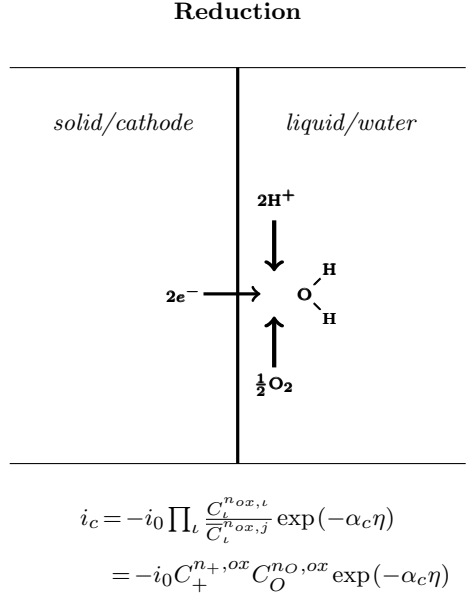
\begin{figure}
\begin{center}
\begin{tikzpicture}[scale=1.5]

\draw (7,4) node {\bf\small Reduction};
\draw[-] (5,3.5) -- (9,3.5) ;
\draw[-] (5,0) -- (9,0) ;
\draw[-, very thick] (7,0) -- (7,3.5);
\draw (6,3) node {\emph{\small solid/cathode}};
\draw (8,3) node {\emph{\small liquid/water}};
\draw (7.7,1.5) node {\setatomsep{1.8em}\pmb{\tiny\chemfig{O(-[1]H)(-[7]H)}}};
\draw[<-,very thick] (7.2,1.5) -- (6.7,1.5); 
\draw (6.53,1.5) node {\pmb{\tiny $2e^-$}};
\draw[<-, ultra thick] (7.32,1.3) -- (7.32,0.85);
\draw (7.35,2.35) node {\pmb{\tiny $2{\rm H}^+$}};
\draw[<-,ultra thick] (7.32,1.7) -- (7.32,2.15); 
\draw (7.32,0.70) node {\pmb{\tiny $\frac{1}{2}{\rm O}_2$}};
\draw (7.0,-0.5) node {\small$i_c=-i_0\prod_\iota\frac{C_\iota^{n_{ox,\iota}}}{\ol{C}_\iota^{n_{ox,j}}}\exp\brkts{-\alpha_c\eta}$};
\draw (7.38,-1) node {\small$=-i_0C_+^{n_+,ox}C_O^{n_O,ox}\exp\brkts{-\alpha_c\eta}$};
\end{tikzpicture}
\caption{
Cathodic current $i_c$ (cathodic branch) defined as a half-cell reaction by a modification of the classical Nernst equation for the activation potential $\Phi$. 
}
\label{fig:BV}
\end{center}
\end{figure}

General electrode reactions can be formulated for reactants $S_\iota$, products $S_j$, and corresponding 
stoichiometric coefficients $\nu_\iota$ and $\nu_j$, respectively, by an abstract equation 
\bsplitl{
\sum_{\iota}\nu_\iota S_\iota +ne^-
		\quad\rightleftharpoons\quad
		\sum_j \nu_jS_j \,,	
}{GeRe}
which can be transformed into an elementary charge transfer reaction as a one-electron reaction
\bsplitl{
\sum_\iota\frac{\nu_\iota}{n}S_\iota + e^-
	\quad\rightleftharpoons\quad
	\sum_j\frac{\nu_j}{n}S_j 
	\,,
}{GER}
where $n$ stands for the number of electrons. By choosing $S_{\iota_1} = O_2$, $\nu_{\iota_1}=\frac{1}{2}$, $S_{\iota_2}=H^+$, $\nu_{\iota_2}=2$, $S_{\iota_3}=e^-$, 
$\nu_{\iota_3}=n=2$, and $S_{j_1}=H_2O$ in \reff{GER}, we recover \reff{R} as the one-electron reaction
\bsplitl{
\frac{1}{4} O_2 + H^+ +e^-
		\quad\rightleftharpoons\quad
		\frac{1}{2}H_2O\,,	
}{oneER}
see Figure \ref{fig:BV}. This allows to define the total electric current density through the pore walls by the Butler-Volmer 
equation
\bsplitl{
i
	= i_0
	\ebrkts{
	\prod_\iota\frac{{C}_\iota^{n_{red,\iota}}}{\ol{C}_\iota^{n_{red,\iota}}}\exp\brkts{\alpha_a\eta}
	-\prod_j\frac{C_j^{n_{ox,j}}}{\ol{C_j}^{n_{ox,j}}}\exp\brkts{-(1-\alpha_a)\eta}
	}\,,
}{TECD} 
where $i_0$ (of order $\ebrkts{\frac{Q}{\tau\ell^2}}$, with $Q$ denoting a reference charge, $\tau$ a characteristic time and $\ell$ a reference length) is the exchange current density defined by
\bsplitl{
\av{i_0}
	:=Fk_a^*\prod_\iota\ol{C}_\iota^{n_{red,\iota}}\exp\brkts{\alpha_a\Phi_0}
	=Fk_c^*\prod_j\ol{C}_j^{n_{ox,j}}\exp\brkts{-(1-\alpha_a)\Phi_0}\,.
}{ExCuDe}
The variable $\eta:=\Phi-\Phi_0$ is the over-potential and the parameter $\alpha_a$ is called anodic charge transfer coefficient. The densities $\ol{C}_\iota$ stand for the bulk concentrations of protons $\iota=+$ and oxygen 
$\iota=O$, respectively, $k^*_a$ and $k^*_c$ are the anodic and cathodic rate constants, 
respectively, and $\Phi_0$ denotes the Nernst equilibrium potential. Further, $n_{red,\iota}$ and $n_{ox,j}$ are the 
reaction orders of the $\iota$-th and $j$-th species, respectively.

It is well-accepted in catalyst layer modeling to take the {\em cathodic} branch of the Butler-Volmer 
equation \reff{TECD}, see \cite{Chan2011}. This approximation is justified at sufficiently large
over-potentials $\eta=\Phi-\Phi_0$. Hence, this branch leads to a reaction rate of  
\bsplitl{
R(C_{+},C_{O},\eta) :=
	i_0C_{+}^{n_{+}}C_{O}^{n_{O}}
	\exp\brkts{-\alpha_c\eta}\,,
}{CaBrBV}
for the cathodic transfer coefficient $\alpha_c:=1-\alpha_a$, the dimensionless concentrations $C_+$ and $C_O$, and 
the reaction orders $n_+=n_{+,ox}$ and $n_O=n_{O,ox}$. 
Note that we follow Chan and Eikerling \cite{Chan2011} regarding the
over-potential $\eta$ which assumes, in fact, only non-positive values. We remark that in electrochemistry one 
often calls $\Phi$ an activation over-potential in difference to the equilibrium potential $\Phi_0$.

\medskip

Let us summarize the relevant equations and refer the reader to Figure \ref{fig:CaPo} for 
convenience:
\begin{center}
{\bf Bulk equations for a single water-filled pore $\Omega^p$} 
\end{center}
\begin{flushleft}
\emph{Oxygen transport:}
\end{flushleft}
\bsplitl{
	-\Div\,\brkts{{\rm D}^O\nabla C_O} = 0
	&\qquad\textrm{in }\Omega^p\,.
}{OT}
\emph{Proton transport:}
\bsplitl{
	-\Div\,\brkts{{\rm D}^+\brkts{\nabla C_+ +C_+\nabla\Phi}} = 0
	&\qquad\textrm{in }\Omega^p\,.
}{PT}
\emph{Electric potential $\phi$:}
\bsplitl{
	-\lambda^2\Delta\Phi = C_+
	&\qquad\textrm{in }\Omega^p\,.
}{EP}

\medskip
\begin{center}
{\bf Boundary conditions on the pore walls $\partial\Omega^p=E_O\cup E_+\cup S^c$}
\end{center}
\begin{flushleft}
\emph{Right entrance of the pore $E_{O}$:}
 \end{flushleft}
\bsplitl{
	C_O = C^D_O
	&\qquad\textrm{on }E_{O},
\\
	-\nabla_nC_+-C_+\nabla_n\Phi = 0
	&\qquad\textrm{on }E_{O}\,,
\\
	\Phi = \Phi^D_{O}
	&\qquad\textrm{on }E_{O}\,,
}{BCL}

\begin{flushleft}\emph{Left entrance of the pore $E_{+}$:} \end{flushleft}
\bsplitl{
-\nabla_nC_O = 0
	&\qquad\textrm{on }E_{+}\,,
\\
C_+ = C_+^D
	&\qquad\textrm{on }E_{+}\,,
\\
\Phi = \Phi^D_H
	&\qquad\textrm{on }E_{+}\,,
}{BCR}

\begin{flushleft}\emph{Pore wall $S^c$:} \end{flushleft}  
\bsplitl{
	-\nabla_nC_O = \frac{1}{4}R(C_{+},C_{O},\eta)
	&\qquad\textrm{on }S^c\,,
\\
	-\nabla_nC_+
	-C_+\nabla_n\Phi
	= R(C_{+},C_{O},\eta)
	&\qquad\textrm{on }S^c\,,
\\
-\epsilon_p\nabla_n\Phi
	= \sigma_s(x)
	&\qquad\textrm{on }S^c\,.
 }{BCW} 
 
\medskip

The surface charge density $\sigma_s$ is introduced in \reff{MaEfEq}. The gradient $\nabla_n:={\bf n}\cdot\nabla$ is defined with respect to a normal vector ${\bf n}$ pointing outward of the pore domain $\partial\Omega^p$. We see that all protons that 
enter the domain are consumed at the wall while water is being 
produced. The latter process is neglected in this contribution so as to simplify the model. 
Strictly speaking, water flow needs to be included as an advective flux. 
 
 \subsection{Microscopic, periodic catalyst layer}\label{sec:MoRe}
\begin{figure}
	\centering
\includegraphics[height=7cm,width=13cm]{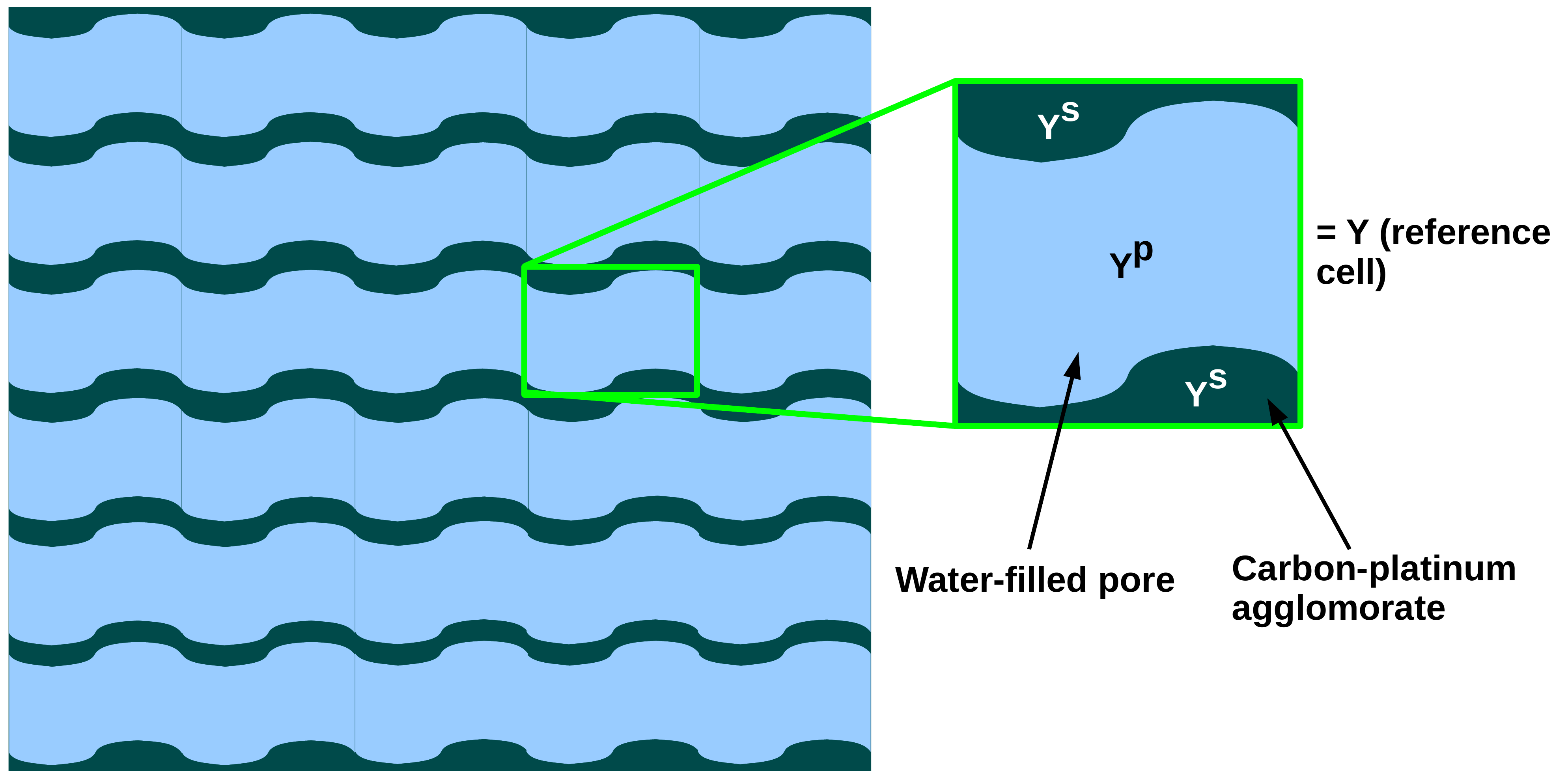} 
	\caption{{\bf Left:} Periodic array (with periodicity scaled by $r$) of water-filled pores $\Omega^p$ between thin catalytic layers $\Omega^s$ containing carbon supported platinum particles. {\bf Right:} Reference cell characterizing the pore geometry.}
	\label{fig:PePo}
\end{figure}

The porous catalyst layer $\Omega$ of characteristic length $L$ decomposes into a pore region $\Omega^p$ and a material domain $\Omega^s$, which represents carbon-supported platinum. The boundary 
$\partial\Omega^p=E_O\cup E_+\cup {\cal I}$ where $E_+$ denotes the left entrance (Fig. \ref{fig:CaPo}), $E_O$ the right entrance (Fig. \ref{fig:CaPo}), and ${\cal I}:=\partial\Omega^p\cap\partial\Omega^s$ the water/solid interface. Furthermore, 
we assume that the pores are periodically distributed in $\Omega$, see Figure \ref{fig:PePo}. The heterogeneity 
of the periodic catalyst layer is defined by the parameter $r:=\frac{\ell}{L}$. We set $\ell=1$ and hence define a periodic unit pore $Y:=[0,\ell]^d$.
The index set
\bsplitl{
K_r := \brcs{j\in\mathbb{Z}^d\,\Bigr|\,{\rm dist}\brkts{rj,\partial\Omega}<r\sqrt{d}}\,,
}{IS}
allows us to denote this pore distribution and the solid regions by
\bsplitl{
\Omega^p_r 
	& := \Omega \cap \bigcup_{z\in\mathbb{Z}^d\setminus K_r}r\brcs{z+Y^p}\,,
\\
\Omega^s_r 
	&:= \Omega \cap \bigcup_{z\in\mathbb{Z}^d\setminus K_r}r\brcs{z+Y^s}\,,
}{Doms}
respectively, where $Y^p\subset Y$ denotes the pore phase in the reference cell $Y$ and 
$Y^s\subset Y$ denotes the heterogeneous C-Pt (carbon nanotubes, carbon meso-pores) 
or polymer-Pt (3M's UTCL) phase. In fact, $Y^p$ 
represents the single pore considered in the previous Section \ref{sec:SiPoMo}.
Note that $\Omega^p$ differs from the periodic replacement $\Omega^p_r$ which is scaled by $r$.
These conventions and assumptions enable us to rewrite the system \reff{OT}-\reff{BCW} in the following dimensionless 
form
\bsplitl{
(\textrm{\bf Micro bulk model:})\quad
\begin{cases}
-\Delta C_O^r = 0
	&\qquad\textrm{in }\Omega_r^p\,,
\\
-\Delta C_+^r - \Div\brkts{C^r_+\nabla\Phi^r} = 0
	&\qquad\textrm{in }\Omega^p_r\,,
\\
-\Div\brkts{{\varepsilon}\brkts{\frac{x}{r}}\nabla\Phi^r} = C_+^r
	&\qquad\textrm{in }\Omega\,,
\end{cases}
}{PeCaLa}
where $\varepsilon(x/r):=\lambda^2\chi_{\Omega^p_r}(x/r)+\gamma\chi_{\Omega^s_r}(x/r)$ is $Y$-periodic, and $C_0^r,\, C_+^r$, and 
$\Phi^r$ satisfy the same boundary conditions on $E_+$ and $E_O$ as imposed in 
\reff{BCL} and \reff{BCR}.  

On the interface ${\cal I}^r:=\partial\Omega^p_r\cap\partial\Omega_r^s$, which is denoted by $S^c$ in Figure \ref{fig:CaPo} 
for the case of a single pore, we have the following boundary conditions
\bsplitl{
(\textrm{\bf Micro interface:})\quad 
\begin{cases}
-\nabla_n C^r_O 
	= r\beta_{O}(C_{O}^r)^{n_{O}}(C_{+}^r)^{n_{+}} {\rm exp}\ebrkts{-\alpha_c\brkts{\Phi^r-\Phi_0}}
	&\qquad\textrm{on }{\cal I}^r\,,
\\
-\nabla_n C_+^r - C_+^r\nabla_n\Phi^r
	= r\beta_{+}(C_{O}^r)^{n_{O}} (C_{+}^r)^{n_{+}} {\rm exp}\ebrkts{-\alpha_c\brkts{\Phi^r-\Phi_0}}
	&\qquad\textrm{on }{\cal I}^r\,,
\\
-{\varepsilon}(x/r)\nabla_n \Phi^r
	= r\sigma_s(x,x/r)
	&\qquad\textrm{on }{\cal I}^r\,,
\end{cases}
}{Ir}
for the parameters $\beta_{O}:=\frac{i_0 L}{4e{\rm D}_O}$ and 
$\beta_{+}:=\frac{i_0 L}{e{\rm D}_+}$ both of dimension $[L^{-1}]$. We emphasize that the $(d-1)$-dimensional volume of the perforation surface 
${\cal I}^r$ increases without bounds in the limit $r\to 0$. Therefore, problems with the boundary conditions 
on the perforation may show degeneration or 
unbounded growth of the solutions (depending on the sign of the coefficient in the Fourier condition). However, 
this phenomenon does not occur if the coefficient of the boundary operator asymptotically vanishes as the 
small parameter tends to zero, or has zero average over the perforation surface. Hence, we scale the right hand-sides in \reff{Ir} by $r>0$.

%% file: mainres.tex
 Before we state the main result in this article, we need to introduce the following 
 definition.
 
\begin{definition}\label{def:LoEq}\emph{(Local equilibrium)} We say that the periodic reference cells $Y$ 
are in local thermodynamic equilibrium if and only if
\bsplitl{
\mu_0^+ = {\rm log}\,C_+(x) +\Phi(x)
	= {\rm const.}
	\qquad\textrm{in }x/r=:y\in Y\,,
}{LoEq}
where $\mu_0^+$ can only assume different values in different reference cells $Y$.
\end{definition}

Subsequently, we employ the notation
\bsplitl{
{\cg M}_{U}(f)
	:=\frac{1}{\av{U}}\int_{U} f \,dx\,,
}{MU}
for arbitray $U\subset\mathbb{R}^d$ and summarize the main result of this article.

\medskip
  
\begin{theorem}\label{thm:EffMo}
 We assume that the periodic reference cells $Y:=Y^p\cup Y^s\subset\mathbb{R}^d$ are in local thermodynamic equilibrium, see Definition 
 \ref{def:LoEq}. Then, the microscopic problem \reff{PeCaLa}--\reff{Ir} admits, for the formal asymptotic 
 expansions
 \bsplitl{
 C_O^r(x) 
	& \sim C_O^0(x) +rC_O^1(x,y) +r^2C_O^2(x,y)+\dots\,,
\\
C_+^r(x) 
	& \sim C_+^0(x) +rC_+^1(x,y) +r^2C_+^2(x,y)+\dots\,,
\\
\Phi^r(x) 
	& \sim \Phi^0(x) +r\Phi^1(x,y) +r^2\Phi^2(x,y)+\dots\,,
 }{AsEx}
 the following leading order macroscopic system
 \bsplitl{
 (\textrm{\bf Upscaled model:})\quad
\begin{cases}
-{\rm div}\brkts{\hat{\rm D}^O\nabla C_O}
	= \ol{\beta}_O(C_+)^{n_+}(C_O)^{n_O}\exp\brkts{-\alpha_c(\Phi-\Phi_0)}\,,
	&\qquad\textrm{in }\Omega\,,
\\
-{\rm div}\brkts{\hat{\rm D}^+\nabla C_+ 
	+ C_+\hat{\rm M}^+\nabla\Phi}
	= \ol{\beta}_+(C_+)^{n_+}(C_O)^{n_O}\exp\brkts{-\alpha_c(\Phi-\Phi_0)}\,,
	&\qquad\textrm{in }\Omega\,,
\\
-{\rm div}\brkts{\hat{\varepsilon}(\lambda^2,\gamma)\nabla\Phi}
	=pC_++\rho_s\,,
	&\qquad\textrm{in }\Omega\,,
\end{cases}
 }{EfEqs}
 where $p:=\frac{\av{Y^p}}{\av{Y}}=\frac{\av{\Omega^p}}{\av{\Omega}}$ stands for the porosity. The effective porous media correctors $\hat{\rm D}^+=\brcs{{\rm d}^+_{ik}}_{1\leq i,k\leq d}$\,, 
 $\hat{\rm D}^O=\brcs{{\rm d}^O_{ik}}_{1\leq i,k\leq d}$\,, 
 $\hat{\rm M}^+=\brcs{{\rm m}^+_{ik}}_{1\leq i,k\leq d}$\,, and 
 $\hat{\varepsilon}=\brcs{{\epsilon}^0_{kl}}_{1\leq i,k\leq d}$  are defined for $\iota\in\brcs{O,+}$ 
 by
 \bsplitl{
 {\rm d}^\iota_{ik}(t,x)
	& := \frac{1}{\av{Y}}
	\int_{Y^p}\sum_{j=1}^d\brcs{\delta_{ik}-\delta_{ij}\partial_{y_j}N^{k}_\iota(t,x,y)}\,dy\,,
\\
{\rm m}^+_{ik}
	&:= \frac{1}{\av{Y}}
	\int_{Y^p}\sum_{j=1}^d\brcs{\delta_{ik}-\delta_{ij}\partial_{y_j}N^{k}_\phi(y)}\,dy\,,
\\
\epsilon^0_{ik}
	:&=
	\frac{1}{\av{Y}}
	\int_Y \sum_{j=1}^d
			{\varepsilon}(y)\brkts{
				\delta_{ik}-\delta_{ij}\partial_{y_j}N^{k}_\phi(y)
			}		
	\,dy
	\,,
 }{ReCePr}
and the integrands $N_\iota^k$ for $\iota\in\brcs{+,O,\phi}$ appearing in \reff{ReCePr} solve the 
reference cell problems
\bsplitl{
N_O^k\,:\quad
\begin{cases}
\quad
-{\rm div}_y\brkts{
		\nabla_y(N_O^{k}(y)-y_k)
	}
	=0
	&\textrm{in }Y^p\,,
\\\quad\qquad
{\rm n}_I\cdot\brkts{
		\nabla_y(N_O^{k}(y)-y_k)
	}
	= 0
	&\textrm{on }I:=\partial Y^p\,,
\\\qquad\quad
N_O^k \textrm{ is $Y$-periodic and ${\cg M}_{Y^p}(N_O^j)=0$}\,,
\end{cases}
\\
N_+^k\,:\quad
\begin{cases}
\quad- \Delta_y (N_+^{k}-y_k)
	= -\Delta_y(N_\phi^{k}-y_k)
	& \qquad\textrm{in }Y^p\,,
\\
\qquad\qquad\brkts{-\nabla_y(N_+^{k}-y_k)
	+\nabla_y(N_\phi^{k}-y_k)}\cdot {\rm n}_I
	=0
	&\qquad\textrm{on }I:=\partial Y^p\,,
\\
\qquad\qquad {N}_+^{k}\textrm{ is $Y$-periodic and ${\cg M}_{Y^p}(N_+^{k})=0$}\,,
	&
\end{cases}
\\
N_\phi^k\,:\quad
\begin{cases}
\quad- {\rm div}_y\brkts{\varepsilon(y)\nabla_y(N_\phi^{k}-y_k)}
	= 0
	& \qquad\textrm{in }Y\,,
\\
\qquad\qquad N_\phi^{k}\textrm{ is $Y$-periodic and 
${\cg M}_Y(N_\phi^{k})=0$}\,.
	&
\end{cases}
}{ReCePr2}
The dimensionless parameters $\ol{\beta}_O$ 
and $\ol{\beta}_+$ in \reff{EfEqs} are defined by
\bsplitl{
\ol{\beta}_O
	:= \frac{i_0 L\Lambda}{4 e {\rm D}_O}
\qquad\textrm{and}\qquad
\ol{\beta}_+
	:= \frac{i_0 Ll\Lambda}{e {\rm D}_+}\,,
}{beta0}
respectively, where $\Lambda:=\av{\partial Y^p\cap \partial Y^s}=\av{I}$.
Finally, the surface charge $\sigma_s$ turns, 
through upscaling, into the background charge
\bsplitl{
\rho_s(x) 
	:= \frac{1}{\av{Y}}\int_{\partial Y^p}\sigma_s(x,y)do(y)\,,
}{BaCh0}
where $do$ denotes the $(d-1)$-dimensional surface measure.
\end{theorem}

\medskip
 
 
 \begin{rem}\label{rem:BC} The boundary conditions imposed on the system \reff{EfEqs} are canonical extensions of 
 the boundary conditions from the single pore case \reff{BCL}--\reff{BCW} with no-flux conditions normal to solid walls 
 on $\partial\Omega$.
 \end{rem}
 
 \begin{rem}\label{rem:ErEs}\emph{(Error estimates)}
 Let us emphasize that in \cite{Schmuck2011b} one can find rigorous error estimates of the porous media Poisson-Nernst-Planck problem for no-flux boundary conditions instead of the Butler-Volmer reactions. For such a scenario, which additionally 
 depends on time $t\geq 0$, the following result holds:

\begin{itemize}
\item[] Under suitable regularity assumptions one achieves that for $t,r>0$, and a generic constant $C>0$, the error 
${\bf E}_r:=\tilde{\bf u}_r-({\bf u}_0+r{\bf u}_1)$ between the exact solution $\tilde{\bf u}_r={\bf T}_r{\bf u}_r$ of the periodic formulation and its homogenized approximation ${\bf u}_0$ 
satisfies the following estimates
\bsplitl{
\textrm{\bf (Error estimates)}\qquad
\begin{cases}
\quad
\N{{\rm E}^1_r}{L^2(\Omega)}^2(t)
	+\N{{\rm E}^2_r}{L^2(\Omega)}^2(t)
	\leq 
		C\brkts{
			r
			+r^2
		}t\,,
		&
\\\quad
\N{{\rm E}_r^3}{H^1(\Omega)}^2(t)
	\leq C\brkts{
		r
		+r^2
	}(t
	+1)
	\,,
	&
\end{cases}
}{r123}
where ${\bf E}_r:=[{\rm E}^1_r,{\rm E}^2_r,{\rm E}^3_r]'$, ${\bf u}_0:=[c_+^0,c_-^0,\phi^0]'$, 
and ${\bf u}_r:=[c_+^r,c_-^r,\phi^r]'$.
\end{itemize}
As one can see in \reff{r123}, the time-dependence might impose additional sources for errors in the effective 
approximation. For a proof and more details we refer the interested reader to \cite{Schmuck2011b}.
\end{rem}
 
We briefly verify non-negativity of the proton concentration 
 $C_+$ in \reff{PT}. To this end, we introduce the following
 
 \medskip
 
{\bf Auxiliary problem:} \emph{Let $C_+$ solve
 \bsplitl{
 -{\rm div}\,\brkts{{\rm D}^+\brkts{\nabla C_+ +\ebrkts{C_+}^+\nabla\Phi}} 
 	= 0
	\qquad\textrm{in }\Omega^p\,,
 }{AuPT}
 where we define the regularization $[x]^+:=\sup\brcs{0, x}$ for $x\in\mathbb{R}$.}
 
 \medskip
 
 \begin{lem}\label{lem:NonNeg} 
 \emph{(Non-negativity)} Concentrations $C_+$ of \reff{AuPT} are almost 
 everywhere non-negative.
\end{lem}

This immediately implies the non-negativity of $C_+$ for the original problem \reff{PT}$_2$. A corresponding 
derivation in the more general context of fluid flow can be found in \cite{SCHMUCK2009}.

\begin{proof}
The definitions $[C_+]^+:=\sup\brcs{C_+,0}$ and $[C_+]^-:=\sup\brcs{-C_+,0}$ allow to multiply 
\reff{AuPT} by the test function $\xi=[c_+]^-$ such that after subsequent integration and integration by parts we obtain
\bsplitl{
&\brkts{{\rm D}^+\nabla\brcs{[C_+]^+-[C_+]^-},\nabla [C_+]^-}
	+ \brkts{\ebrkts{[C_+]^+-[C_+]^-}^+{\rm D}^+\nabla\Phi,\nabla [C_+]^-}
	 = 0\,,
}{nn1}
where we took advantage of the definitions $[C_+]^+$ and $[C_+]^-$. Since the terms $\ebrkts{[C_+]^+-[C_+]^-}^+$ 
and $[C_+]^-$ are zero on complementary sets, we end up with
\bsplitl{
	-{\rm D}^+\|\nabla [C_+]^-\|^2_{L^2}
	\geq 0\,.
}{nn2}
Since all the terms in \reff{nn2} are positive, we have $[C_+]^-=0$ a.e.  in $\Omega$ and hence non-negativity 
of solutions of \reff{AuPT}.
\end{proof}

Via the proof of Lemma \ref{lem:NonNeg} it follows immediately that solutions to \reff{EfEqs}$_2$ are 
non-negative if and only if 
\bsplitl{
R_{eff}(C_+,C_O,\eta)
	:=\beta_+(C_+)^{n_+}(C_O)^{n_O}\exp\brkts{-\alpha_c\eta}
	\geq 0
	\,,
}{Reff}
which follows with the non-negativity of $C_O$.

The next result of the article is initiated in the context of porous media in \cite{Schmuck2011}. We consider 
situations where the electrical double layers are thin compared to the mean catalyst layer size $L>0$. 
The thin-double-layer approximation is a well-accepted approximation in electrochemistry and 
 engineering for systems based on the Nernst-Planck equations such as \reff{EfEqs}. This approximation consists mathematically of passing to the limit $\lambda:=\frac{\lambda_D}{L}\to 0$. However, one 
immediately recognizes that such a limit alone does not have much influence on the form of \reff{EfEqs}. Hence, we 
additionally take the limit $\gamma\to 0$, i.e., we assume that the porous matrix is insulating, see also 
\cite{Schmuck2011a}. 

\medskip

\begin{corollary}\emph{(Thin double layers in pores)}\label{lem:ThD}
Let us assume that the solid phase of the catalyst layer forms an insulating matrix and we know the porosity $p$, 
effective diffusion tensors $\hat{\rm D}^O$ and $\hat{\rm D}^+$, and the homogenized surface charge $\rho_s$. Then, the leading order  
bulk approximation for oxygen concentration $C_O$, proton concentration $C_+$, and electric potential $\Phi$ 
reads
\bsplitl{
\begin{cases}
\quad
-{\rm div}\brkts{\hat{\rm D}^O\nabla C_O}
	 = \beta_O\brkts{-\frac{\rho_s}{p}}^{n_+}(C_O)^{n_O}\exp\brkts{-\alpha_c(\Phi-\Phi_0)}\,,
\\
\quad
-{\rm div}\brkts{
		\frac{\rho_s}{p}\hat{\rm D}^+\nabla\Phi
	}
	 = \beta_+\brkts{-\frac{\rho_s}{p}}^{n_+}(C_O)^{n_O}\exp\brkts{-\alpha_c(\Phi-\Phi_0)}
	 -{\rm div}\brkts{
		\hat{\rm D}^+\nabla\frac{\rho_s}{p}  
	 }\,,
\\
\quad
 C_+ = -\frac{\rho_s}{p}\,.
\end{cases}
}{LeOrBuCo}
\end{corollary}

\medskip

We immediately recognize that \reff{LeOrBuCo} is different from thin-double-layer approximations of classical 
Nernst-Planck-Poisson systems since we do not have counter ions $C_-$ leading to the well-known 
quasi-electroneutrality in the electrolyte. Therefore, the electric potential solves now a nonlinear 
Poisson equation due to the upscaled interfacial reactions.

%% file: multiScale.tex
We look for solutions in the form of the two-scale asymptotic expansions
\bsplitl{
C_O^r(x) 
	& \sim C_O^0(x) +rC_O^1(x,y) +r^2C_O^2(x,y)+\dots\,,
\\
C_+^r(x) 
	& \sim C_+^0(x) +rC_+^1(x,y) +r^2C_+^2(x,y)+\dots\,,
\\
\Phi^r(x) 
	& \sim \Phi^0(x) +r\Phi^1(x,y) +r^2\Phi^2(x,y)+\dots\,,
}{AsSe}
where functions are periodic with respect to the microscopic coordinate $y:=x/r$. We emphasize that the ansatz 
\reff{AsSe} is only formal since the convergence is not guaranteed and possible boundary layers are neglected. Moreover, we already took into 
account that the leading order terms are independent of the microscale $y$, see \cite{Schmuck2011,Schmuck2011a} for instance. The 
following identities
\bsplitl{
\frac{\partial \varphi(x,y)}{\partial x_i}
	& = \partial_{x_i}\varphi(x,y) + \frac{1}{r}\partial_{y_i}\varphi(x,y)\,,
\\
\frac{\partial^2\varphi(x,y)}{\partial x^2_i}
	& = \partial^2_{x_i}\varphi(x,y) 
		+ \frac{1}{r}\brkts{\partial_{x_i}\partial_{y_i} +\partial_{y_i}\partial_{x_i}}\varphi(x,y)
		+ \frac{1}{r^2}\partial^2_{y_i}\varphi(x,y)\,,
}{DiOp}
are an immediate consequence of the definition of the small scale variable $y$.
\medskip 
After substituting \reff{AsSe}$_1$ into \reff{PeCaLa}$_1$ and collecting terms of equal power in $r$, we obtain
a recurrent sequence of problems. The first has the form
\bsplitl{
-\sum_{k,j=1}^d\partial_{y_k}\brkts{\delta_{kj}\partial_{y_j}C_O^1} 
	- \sum_{k,j=1}^d\partial_{x_k}\brkts{\delta_{k,j}\partial_{y_j}C_O^0} = 0
	& \qquad\textrm{in }Y^p\,,
\\
-\nabla_{y_n}C_O^1 
	- \nabla_{x_n}C_O^0
	= 0
	&\qquad\textrm{on }{\cal I}^r\cap Y\,,
}{H1}
where $x\in\Omega$ plays the role of a parameter. An integral identity corresponding to problem 
\reff{H1} reads for all $V\in H^1_\#(Y^p)$ 
\bsplitl{
\sum_{k,j=1}^d
	\int_{Y^p}\delta_{kj}\partial_{y_j}C_O^1\partial_{y_k}V\,dy
	& + \sum_{k,j=1}^d\int_{Y^p}\delta_{kj}\partial_{x_j}C_O^0\partial_{y_k}V\,dy
	= 0\,.
}{IId1}
This equation suggests to choose $C_O^1$ as 
\bsplitl{
C_O^1(x,y) 
	= 
	-\sum_{i=1}^dN_O^i(y)\partial_{x_i}C_O^0(x)\,.
}{Sub1}
By inserting \reff{Sub1} into \reff{IId1}, we obtain the following 
problem for $N_O^j(y)$ and $1\leq j\leq d$, i.e.,
\bsplitl{
\partial_{x_j}C_O^0(x)
	\sum_{k,l=1}^d\brcs{\int_{Y^p}\delta_{kl} 
 	\partial_{y_l}N_O^j(y)
	\partial_{y_k}V\,dy
	-\sum_{k=1}^d\int_{Y^p}\delta_{kj}\partial_{y_k}V\,dy}
	= 0
	\,.	
}{NO}
In the classical formulation, \reff{NO} reads
\bsplitl{
\begin{cases}
\quad
-{\rm div}_y\brkts{
		\nabla_y(N_O^{j}(y)-y_j)
	}
	=0
	&\textrm{in }Y^p\,,
\\\quad\qquad
{\rm n}_I\cdot\brkts{
		\nabla_y(N_O^{j}(y)-y_j)
	}
	= 0
	&\textrm{on }I:=\partial Y^p\,,
\\\qquad\quad
N_O^j \textrm{ is $Y$-periodic and ${\cg M}_{Y^p}(N_O^j)=0$}\,.
\end{cases}
}{clNO}
We note that the last line in \reff{clNO} is imposed in order to account for the fact that $N^i_O$ is only 
determined up to an additive constant. The next problem in the recurrent chain has the form
\bsplitl{
-\sum_{k,j=1}^d\partial_{y_k}\brkts{\delta_{kj}\partial_{y_j}C_O^2(x,y)}
	-\sum_{k,j=1}^d\partial_{x_k}\brkts{\delta_{kj}\partial_{y_j}C_O^1(x,y)}
	\quad\qquad\qquad\qquad&
\\	
	-\sum_{k,j=1}^d\partial_{y_k}\brkts{\delta_{kj}\partial_{x_j}C_O^1(x,y)}
	-\sum_{k,j=1}^d\partial_{x_k}\brkts{\delta_{kj}\partial_{x_j}C_O^0(x)}
	= 0\,,
	&\qquad\textrm{in }Y^p\,,
\\
-\nabla_{y_n}C_O^2
	-\nabla_{x_n}C_O^1 
	= \beta_O(C_+^0)^{n_+}(C_O^0)^{n_O}{\rm exp}\ebrkts{-\alpha_c\brkts{\Phi^0-\Phi_0}}\,\,,\,
	&\qquad\textrm{on }{\cal I}^r\cap Y\,.
}{H12}
The integral formulation of \reff{H12} reads
\bsplitl{
&\sum_{k,j=1}^d\int_{Y^p}\delta_{kj}\partial_{y_j}C_O^2\partial_{y_k}V\,dy
	-\sum_{k,j=1}^d\int_{Y^p}\delta_{kj}\partial_{x_k}\brkts{\partial_{y_j}C_O^1}V\,dy
	+\sum_{k,j=1}^d\int_{Y^p}\delta_{kj}\partial_{x_j}C_O^1\partial_{y_k}V\,dy
	\\&
	-\sum_{k,j=1}^d\int_{Y^p}\delta_{kj}\partial_{x_k}\brkts{\partial_{x_j}C_O^0}V\,dy
	-\sum_{k,j=1}^d\int_{{\cal I}^r\cap Y}\beta_O(C_+^0)^{n_+}(C_O^0)^{n_O}{\rm exp}\ebrkts{-\alpha_c\brkts{\Phi^0-\Phi_0}}V\,d\sigma(y)
	= 0\,.
}{VH12}
The second and third term in \reff{VH12} can be rewritten by \reff{Sub1} as
\bsplitl{
-\sum_{k,j=1}^d\int_{Y^p}\delta_{kj}\partial_{y_j}C_O^1\partial_{x_k}V\,dy
	& = -\sum_{k,j=1}^d\int_{Y^p}\delta_{kj}\partial_{y_j}N_O^i(y)\partial_{x_i}C_O^0(x)\partial_{x_k}V\,dy\,,
\\
\sum_{k,j=1}^d\int_{Y^p}\delta_{kj}\partial_{x_j}C_O^1\partial_{y_k}V\,dy
	& = \sum_{k,j=1}^d\int_{Y^p}\delta_{kj}N_O^i(y)\partial_{x_i}\partial_{x_j}C_O^0(x)\partial_{y_k}V\,dy\,.
}{RW1}
The solvability condition for problem \reff{VH12} is an equation for $C_O^0(x)$. It is the required homogenized 
(limit) equation. By defining 
\bsplitl{
{\rm d}^O_{ik} 
	& := \sum_{j=1}^d\int_{Y^p}\brkts{-\delta_{ij}\frac{\partial N^k_O(y)}{\partial y_j}+\delta_{ik}}\,dy\,,
\\
\Lambda 
	& := \av{\partial Y^p\cap \partial Y^s}\,,
}{Ds}
and after setting $V\equiv 1$ in \reff{VH12}, we end up with the homogenized equation
\bsplitl{
-\sum_{k,j=1}^d\partial_{x_k}\brkts{{\rm d}^O_{kj}\partial_{x_j}C_O^0} 
	= \ol{\beta}_O(C_+^0)^{n_+} (C_O^0)^{n_O}{\rm exp}\ebrkts{-\alpha_c\brkts{\Phi^0-\Phi_0}}\,,
}{OxTr}
where $\ol{\beta}_O:=\Lambda\beta_O$ is dimensionless with $\beta_O$ defined after \reff{Ir}.

\medskip

The important information gained by the calculation of \reff{OxTr} is how the boundary condition 
\reff{Ir}$_1$ enters in the upscaled or homogenized problem. Since the equation \reff{PeCaLa}$_2$ satisfies the 
same type of boundary condition on the interface as \reff{PeCaLa}$_1$, we take over 
the result from \reff{OxTr}. Moreover, in view of the two-scale convergence analysis from \cite{Schmuck2010}, the macroscopic equations for the remaining two equations in \reff{PeCaLa} immediately turn into
\bsplitl{
\begin{cases}
\quad
-{\rm div}\brkts{\hat{\rm D}^+\nabla C^0_++C^0_+\hat{\rm M}^+\nabla\Phi^0}
	= \ol{\beta}_+(C_+^0)^{n_+}(C_O^0)^{n_O}{\rm exp}\ebrkts{-\alpha_c\brkts{\Phi^0-\Phi_0}}
	&\qquad\textrm{in }\Omega\,,
\\
\quad
-{\rm div}\brkts{\hat{\varepsilon}\brkts{\lambda^2,\gamma}\nabla\Phi^0}
	= p C_+^0
	&\qquad\textrm{in }\Omega\,,
\end{cases}
}{PrTrPhi}
where $\ol{\beta}_+:=\Lambda\beta_+$ for $\beta_+$ defined after \reff{Ir}. Further, the correction tensors $\hat{\rm D}^\iota:=\brcs{{\rm d}^\iota_{ik}}_{1\leq i,k\leq d}$, $\hat{\rm M}^+:=\brcs{{\rm m}^+_{ik}}_{1\leq i,k\leq d}$, and 
$\hat{\varepsilon}:= \brcs{\varepsilon^0_{ik}}_{1\leq i,k\leq d}$ are defined by
\bsplitl{
{\rm d}^\iota_{ik}
	:= \frac{1}{\av{Y}}
	\sum_{j=1}^d\int_{Y^p}\brcs{\delta_{ik}-\delta_{ij}\partial_{y_j}N_\iota^{k}(y)}\,dy
	\qquad\forall i,k=1,\dots,N\,,
\\
{\rm M}^+_{ik}
	:= \frac{1}{\av{Y}}
	\sum_{j=1}^d\int_{Y^p}\brcs{\delta_{ik}-\delta_{ij}\partial_{y_j}N_\phi^{k}(y)}\,dy
	\qquad\forall i,k=1,\dots,N\,,
\\
\epsilon^0_{ik}
	:=
	-\frac{1}{\av{Y}}
	\sum_{j=1}^d\int_Y
		\brcs{
			\varepsilon(y)\brkts{
				\delta_{ik}-\delta_{ij}\partial_{y_j}N_\phi^{k}
			}
		}
	\,dy
	\qquad\forall i,k=1,\dots,N\,,	
}{DandE}
for $\gamma :=\frac{\varepsilon_s}{\varepsilon_p}$, $\lambda := \sqrt{\frac{\varepsilon_pRT}{2z_+^2eFC_0}}/L$, 
and $\iota\in\brcs{+,O}$. The correctors $N^k_+$ and $N^k_\phi$ solve the following reference cell problems
\bsplitl{
\begin{cases}
\quad- \Delta_y (N_+^{k}-y_k)
	= -\Delta_y(N_\phi^{k}-y_k)
	& \qquad\textrm{in }Y^p\,,
\\
\qquad\qquad\brkts{-\nabla_y(N_+^{k}-y_k)
	+\nabla_y(N_\phi^{k}-y_k)}\cdot {\rm n}_I
	=0
	&\qquad\textrm{on }I:=\partial Y^p\,,
\\
\qquad\qquad {N}_+^{k}\textrm{ is $Y$-periodic and ${\cg M}_{Y^p}(N_+^{k})=0$}\,,
	&
\\
\quad- {\rm div}_y\brkts{\varepsilon(y)\nabla_y(N_\phi^{k}-y_k)}
	= 0
	& \qquad\textrm{in }Y\,,
\\
\qquad\qquad N_\phi^{k}\textrm{ is $Y$-periodic and 
${\cg M}_Y(N_\phi^{k})=0$}\,.
	&
\end{cases}
}{DiCoEq}
The reference cell problems \reff{clNO}, \reff{DandE}, and \reff{DiCoEq}, written in classical form, are mathematically only meaningful in the distributional sense. Existence and uniqueness follows then by Lax-Milgram's theorem 
for a suitable weak formulation. For rigorous solvability results we refer the interested reader to \cite{Schmuck2011,Schmuck2010,Schmuck2011a}. As a consequence, it is suggested to apply Galerkin schemes (e.g. finite elements) for the computation of numerical solutions.

\medskip

We note that \reff{MaEfEq}, which is obtained from the results in this section by dropping the superscript ``0'' 
in \reff{OxTr} and \reff{PrTrPhi}, can immediately be extended to account for a surface charge density $\sigma_s$ 
on the pore walls denoted by ${\cal I}$. Such a density enters on the right hand side in the third equation of the effective model \reff{MaEfEq} as the background charge
\bsplitl{
\rho_s(x) 
	:= \frac{1}{\av{\partial Y^p}}\int_{Y}\sigma_s(x,y)do(y)\,,
}{BaCh}
see \cite{Allaire1996,Schmuck2011} for a derivation. We promote this idea as a straightforward way to account for double layer 
effects without going into details of Stern or Helmholtz layers, see \cite{Bard2001,Plieth2008} for an overview of such models and \cite{Biesheuvel2011,Chan2011} for applications. Here, the use of a surface charge is appropriate since \reff{MaEfEq} is time-independent and hence models the steady case where the 
double layers are in local thermodynamic equilibrium.

In future work, we will apply our algorithm and methodology to a number of examples, especially
3M's nano-structured thin-film catalyst layers.

%% file: applic.tex
\begin{figure}

\center

\begin{tikzpicture}[scale=1]

	\setlength{\unitlength}{1cm}
		
	\foreach \x in {1,...,6}
		\foreach \y in {1,...,5}
		{
		\color{grau}
			\fill (\x,\y) rectangle +(-1,-.2);
			\fill (\x,\y-.8) rectangle +(-1,-.2);
			\pgfsetlinewidth{1pt};\color{red}
			\draw[-] (\x,\y-0.2) -- (\x-1,\y-0.2);
			\draw[-] (\x,\y-0.8) -- (\x-1,\y-0.8);	
		\pgfsetlinewidth{2pt}; \color{black}
			\draw (\x,\y) rectangle (\x-1,\y-1);
		
		}

	\pgfsetlinewidth{2pt};
	\Large
	
	\draw[-] (6,3) -- (6.5,3);
	\draw[-] (6,4) -- (6.5,4);
	\pgfsetlinewidth{1pt};
	\draw[->] (6.25,3.5) -- (6.25,4);
	\draw[->] (6.25,3.5) -- (6.25,3);
	\put(6.33,3.35){\color{blue}$r$};

	\pgfsetlinewidth{2pt};	
	\draw[->] (-2,0) -- (-1,0);
	\draw[->] (-2,-0.035) -- (-2,1);
	\put(-0.9,-0.08){\color{blue}$x_1$};
	\put(-2.2,1.15){\color{blue}$x_2$};

	\color{red}
	\put(-2,3.5){$\rho_s$};
	\pgfsetlinewidth{2pt};
	\draw[->](-1.5,3.5) -- (1.5,2.22);	
		
\end{tikzpicture}
\caption{A periodic catalyst layer defined by a single reference cell $Y:=[0,1]^2$ scaled by $r$. The variable 
$\rho_s$ denotes the effective surface charge at the pore-solid interface. $x_3$ is pointing out of the plane.} 
\label{fig:StCh}
\end{figure}
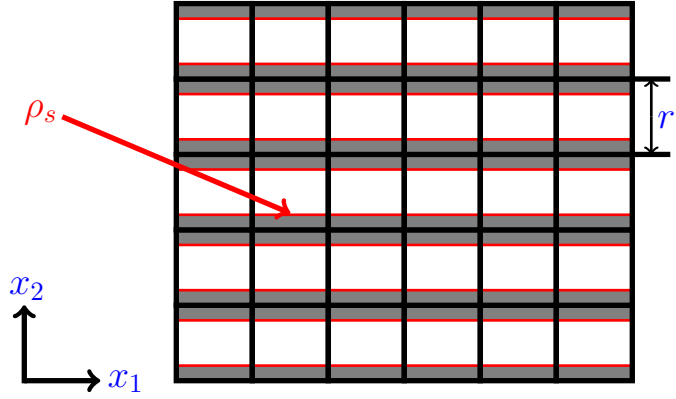

Auriault and Lewandowska \cite{Auriault1997} considered the situation of a porous medium 
defined by a periodic array of straight channels as depicted in Figure \ref{fig:StCh}. They analytically 
compute a correction tensor in order to define a homogenized diffusion coefficient in porous structures. 
We make use of their results herein and hence restrict ourselves to the case of an insulating 
porous matrix, i.e., $\gamma\to 0$. Then, we obtain
\bsplitl{
\hat{\rm D}^O
	= \hat{\rm M}^+
	= \ebrkts{\begin{matrix}
	p & 0 & 0
\\
	0 & 0 & 0
\\
	0 & 0 & p
	\end{matrix}}
\qquad\textrm{and}\qquad
\hat{\varepsilon}
	= \ebrkts{\begin{matrix}
	p\lambda^2 & 0 & 0
\\
	0 & 0 & 0
\\
	0 & 0 & p\lambda^2
	\end{matrix}}
	\,,
}{StCh}
where $p$ denotes the porosity and the zero-entry on the diagonal is the direction 
orthogonal to the direction of the channel, see Figure \ref{fig:StCh}. Under \reff{StCh} the macroscopic 
catalyst layer model becomes in the two-dimensional case
\bsplitl{
(\textrm{\bf Straight channel:})\quad
\begin{cases}
-p\brkts{
		\frac{\partial^2}{\partial x_1^2} 
		+\frac{\partial^2}{\partial x_3^2}
	}C_O
	= \ol{\beta}_O(C_+)^{n_+}(C_O)^{n_O}\exp\brkts{-\alpha_c(\Phi-\Phi_0)}\,,
	&\qquad\textrm{in }\Omega\,,
\\
-p\brkts{
		\frac{\partial^2}{\partial x_1^2} 
		+\frac{\partial^2}{\partial x_3^2}
	} C_+
	-p\frac{\partial}{\partial x_1}\brkts{
		C_+
		\frac{\partial}{\partial x_1}\Phi}
	-p\frac{\partial}{\partial x_3}\brkts{
		C_+
	\frac{\partial}{\partial x_3}\Phi}
\\
	\qquad\qquad\qquad\qquad\,\,\,
	= \ol{\beta}_+(C_+)^{n_+}(C_O)^{n_O}\exp\brkts{-\alpha_c(\Phi-\Phi_0)}\,,
	&\qquad\textrm{in }\Omega\,,
\\
-p\lambda^2\brkts{
		\frac{\partial^2}{\partial x_1^2} 
		+\frac{\partial^2}{\partial x_3^2}
	}\Phi
	=pC_++\rho_s\,,
	&\qquad\textrm{in }\Omega\,.
\end{cases}
}{CLstCh}
The coordinate $x_2$ acts like a parameter in the system \reff{CLstCh}. We recover the 
results from \cite{Auriault1997} if we set $\Phi=0$ and $\ol{\beta}_+=\ol{\beta}_O=0$.

Next, we want to apply the scenario of straight channels to the thin-double-layer 
approximation \reff{LeOrBuCo}. As in \reff{CLstCh}, we consider an insulating 
porous matrix. We immediately end up with
\bsplitl{
\begin{cases}
\quad
-p\brkts{\frac{\partial^2}{\partial x_1^2}
	+\frac{\partial^2}{\partial x_3^2}
	}C_O
	 = \ol{\beta}_O\brkts{-\frac{\rho_s}{p}}^{n_+}(C_O)^{n_O}\exp\brkts{-\alpha_c(\Phi-\Phi_0)}\,,
\\
\quad
-\frac{\partial}{\partial x_1}\brkts{
		\rho_s\frac{\partial}{\partial x_1}\Phi
	}
	-\frac{\partial}{\partial x_3}\brkts{
		\rho_s\frac{\partial}{\partial x_3}\Phi
	}
	 = \ol{\beta}_+\brkts{-\frac{\rho_s}{p}}^{n_+}(C_O)^{n_O}\exp\brkts{-\alpha_c(\Phi-\Phi_0)}
	 -\brkts{\frac{\partial^2}{\partial x_1^2} +\frac{\partial^2}{\partial x_3^2}}\rho_s
	\,,
\\
\quad
 C_+ = -\frac{\rho_s}{p}\,,
\end{cases}
}{LeOrBu}
where the coordinate $x_2$ acts again like a parameter. It is interesting to see that the porosity 
parameter $p$ cancels out in \reff{LeOrBu}$_2$ except for the reaction term.

In future work, we will extend the upscaled equations towards fluid flow and compute effective transport coefficients 
based on the formulas \reff{ReCePr} and \reff{ReCePr2} for carbon nanotubes \cite{Simon2008} and UTCLs such as 3M's organic perylene whiskers \cite{Debe2006}.